\UseRawInputEncoding
\documentclass{article}
\usepackage{graphicx} 
\usepackage{CJKutf8}
\usepackage[utf8]{inputenc}
\usepackage{amsmath}
\usepackage{indentfirst}
\usepackage{caption}
\usepackage{cite}
\usepackage{hyperref}
\usepackage{authblk}
\hypersetup{
hypertex=true,
colorlinks=true,
linkcolor=blue,
anchorcolor=blue,
citecolor=blue}

\title{Quantum erasure based on phase structure}
\author[1]{\small Ye Yang}
\author[1]{Chengyuan Wang}
\author[2]{Yun Chen}
\author[1]{Jianyi Xv}
\author[1]{Xin Yang}
\author[1]{Jinwen Wang}
\author[1]{Shuwei Qiu}
\author[1, $\dagger$]{Hong Gao}
\author[1]{Fuli Li}

\affil[1]{MOE Key Laboratory for Nonequilibrium Synthesis and Modulation of Condensed Matter, Shaanxi
Province Key Laboratory of Quantum Information and Quantum Optoelectronic Devices, School of Physics,
Xi‘an Jiaotong University, 710049, China}
\affil[2]{Department of Physics, Huzhou University, Huzhou 313000, China}
\affil[$\dagger$]{Email: honggao@xjtu.edu.en}

\date{May 2024}

\begin{document}
\begin{CJK}{UTF8}{gbsn}
\maketitle

\begin{abstract}
The quantum eraser effect exemplifies the distinct properties of quantum mechanics that challenge classical intuition and expose the wave-particle duality of light.
This effect has been extensively explored in various experiments; most of these investigations use polarisation to distinguish which path information, and less attention has been paid to the phase structure which is related wavefront of photon. In this study, we introduce a theoretical framework for quantum erasure that focusses on the phase structure and demonstrate it experimentally. 
 In this experiment, we employ a Mach-Zehnder interferometer (MZI) where a first-order spiral phase plate (SPP) is integrated into one of its arms. This setup applied orbital angular momentum (OAM) to the photons and established predetermined which-way information. Consequently, the photon demonstrates its particle characteristics, with absence of interference at the MZI's output ports. Utilizing an additional SPP to erase the phase structure from the output photon results in pronounced interference patterns, observable in a post-measurement scenario. This result allows us
to include the structure information of the equiphase plane of the light field in quantum erasure. The results challenge the traditional cause-effect relationship in classical physics, given that the subsequent choice of the SPP adheres to a space-like separation.
\end{abstract}

\section{Introduction}

\par In 1928, Bohr proposed the "complementarity principle" to explain the "wave-particle duality" of light\cite{Bohr}. This principle suggests that whether a photon exhibits wave-like or particle-like behaviour depends on whether the path information could be discriminated. Extensive research has been conducted in order to fully comprehend and elucidate this fascinating phenomenon, with two of the most prominent studies being the quantum delayed choice (QDC) and the quantum eraser (QE). The QDC experiment was first conceived by Wheeler\cite{WHEELER19789}, in which the choice of particle measurement or interferometry is made after the photon has already entered the interferometer to rule out the possibility of predicting which measurement it will encounter. A further assumption is that if a photon carrying path information were to leave the interferometer and subsequently erase its path information, what would be the result? Soon after, Scully and Drühl proposed that a QE can erase the which-path information even after the quantum itself has left the interferometer and determine its early behaviour as wave-like or particle-like\cite{Scully1982,Scully1991}. Since then, several QE experiments have been conducted\cite{Herzog1995,Walborn2002,Kim2000,Wheeler1983QuantumTA,Mohrhoff1996,Mohrhoff1999,Englert1999}. Such fundamental experiments have not only realized historic proposals but have also helped sharpen our understanding of wave-particle duality.

\par The QDC and QE experiments can be conducted using entangled photons\cite{Kim2000,Kaiser2012,RN370}, thermal light\cite{RN304}, single photons\cite{RN374,RN389}, attenuated lasers\cite{RN309}, and other particles such as atom\cite{RN312,RN362} and electron\cite{RN381}. The ongoing advancement of these progressive steps is gradually expanding the scope of human perception to encompass the quantum realm. When adopting photons to investigate the QE phenomenon, the polarization basis is commonly employed to facilitate path information identification due to its flexible modulation capability \cite{Walborn2002,Jacques2007,Jacques2008,Kaiser2012,Peruzzo2012,Tang2012,Nape2017,RN309}. In addition to polarization, the phase distribution of photons also plays a crucial role in determining the interference outcomes. However, up to now, employing the photonic phase structure to study QE remains unexplored. 

\par In this paper, we provides a theoretical model and experimental validation of quantum erasure based on the phase structure of light. Specifically, we construct a typical Mach–Zehnder interferometer (MZI) with a first-order spiral phase plate (SPP) inserted into one arm of the MZI, which applies an orbital angular momentum (OAM)\cite{RN444} to the photon and pre-determines the which-way information of the photon. Hence the photon exhibits its particle nature with no interference observed in the output ports of the MZI. When we use another SPP to erase the phase information of the output photon, high-contrast interference occurs in such a post-measurement manner. These outcomes violate the cause-effect relation in classical physics, as the postchoice of the SPP satisfies the space-like separation.

\section{Theoretical Analysis}
\begin{figure}
    \centering
    \includegraphics[width=0.75\linewidth]{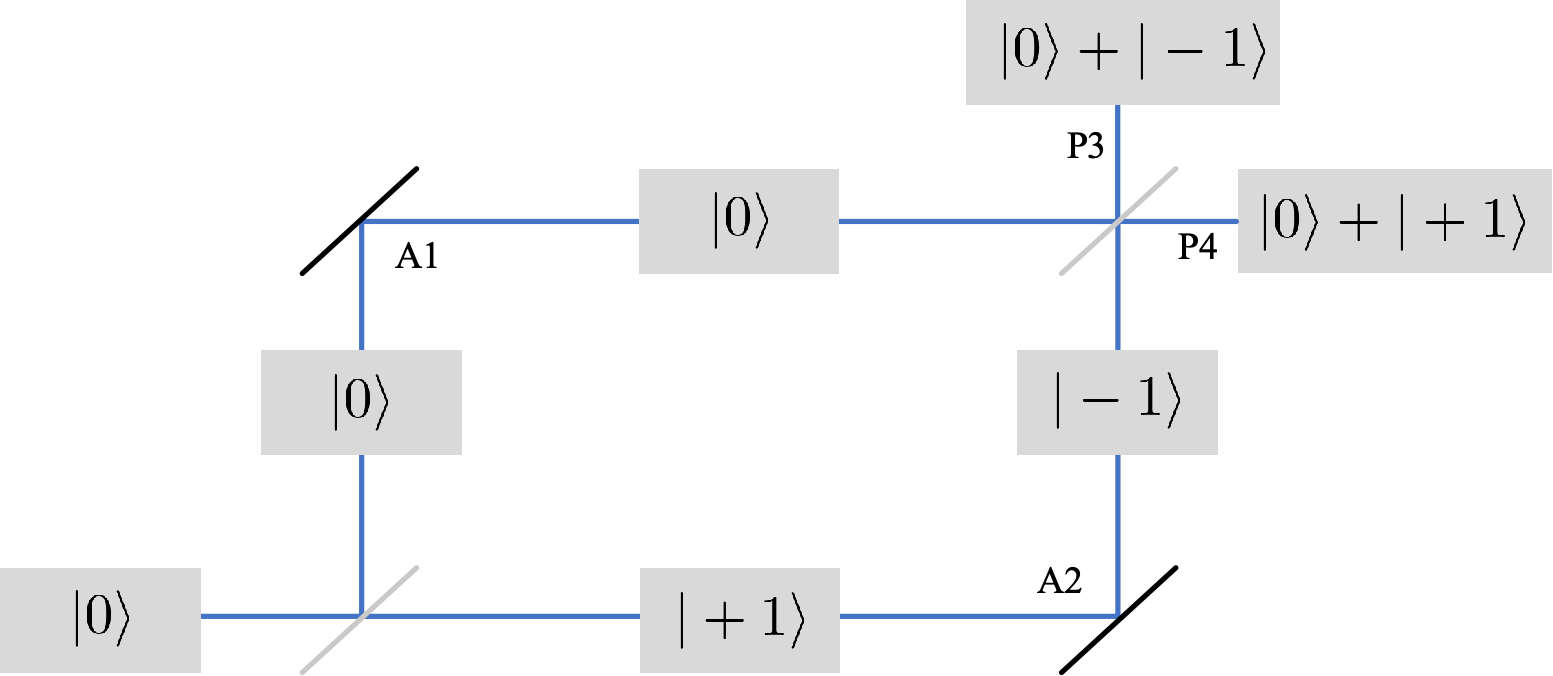}
    \caption{Logic diagram of the experiment consisting of two BS, two mirrors and one SPP ( at A2 but not shown). A1, upper arm of MZI; A2, lower arm of MZI; P3, upper path of output; P4, lower path of output.}
    \label{Fig1}
\end{figure}
The quantum erasure experiment hinges on whether to eliminate photon path information. If the propagation path of the photon is determined, then the photon will exhibit its particle-like behaviour without interference. However, if the path information is erased, the photon will demonstrate its wave-like behavior and interfere. The structure of the phase of light refers to the phase distribution of the light field in its propagation cross section, and can also be considered as its equiphase surface structure in the direction of propagation. A photon-carrying OAM can be prepared by constructing its phase structure. The most typical equiphase surface is the spiral described by $LG^{0}_{1}(r,\theta,\phi)$\cite{RN444}. To enhance the natural and intuitive description, the basis vector $|l \rangle$ in the OAM representation is used to describe the quantum state of the light field, where $l$ is the quantum number of OAM and the OAM carried by each photon in the state $|l \rangle$ is $l\hbar$.
The operator $\hat{S}$ defines the effect of the spiral phase plate (SPP) on the light field. The $\hat{S}^{\dagger}$ of the SPP corresponds to the direction in which the quantum number of the light field $l$ increases, while the $\hat{S}$ corresponds to the direction in which it decreases. Therefore, the relation $\hat{S}|l\rangle=e^{i\theta}|l-1\rangle$, $\hat{S}^{\dagger}|l\rangle=e^{-i\theta}|l+1\rangle$ holds, where the phase of the SPP is indicated as $\theta$.

\par To investigate the OAM-based quantum eraser experiment, we constructed a Mach–Zehnder interferometer (MZI) with a Gaussian beam attenuated to the single-photon level as input. The two arms of the MZI are labelled A1 and A2. A first-order spiral phase plate (SPP) is inserted in arm A2, which can change the phase structure of the beam passing through it. For instance, when a beam carrying an OAM with a quantum number of $l$ (the OAM state is denoted as $|l\rangle$) passes through the center of the SPP, the state will become $\hat{S}|l\rangle=|l+1\rangle$, where the operator $\hat{S}$ corresponds to the function of the SPP. Hence, the Gaussian beam passing through the centre of the SPP becomes $\hat{S}|0\rangle=|1\rangle$.
Figure \ref{Fig1} illustrates that when the OAM beam is reflected by a mirror (M) or beam splitter (BS), the sign of the OAM state is inverted. The MZI's output ports (P3 and P4) hold the superposition of the states in the two arms. The optical states are described as follows:
\begin{align*}
|\psi_{P3}\rangle=\frac{1}{\sqrt{2}}(|0\rangle+e^{i\phi}|-1\rangle)\tag{1}\\
        |\psi_{P4}\rangle=\frac{1}{\sqrt{2}}(|0\rangle+e^{i\phi}|1\rangle)\tag{2}
\end{align*}

where $\phi$ is the phase difference between the two arms. If $|0\rangle$ mode is detected by a detector placed at P3 or P4, we will know that the photon comes from the A1 arm. Otherwise, if the $|+1\rangle$ or $|-1\rangle$ mode is detected, the photon comes from the A2 arm. In this situation, the photon exhibits 
particle-like behavior since its path is distinguishable. 

Now, let us consider another case. A first-order SPP is placed at P3 and a first-order SPP is placed at P4, and the centre of the SPP is shifted by a distance of $r/2$ with respect to the centre of the Gaussian beam ($r$ is the radius of the Gaussian beam). The operations corresponding to these SPPs are described below: 
\begin{align*}
    \hat{S}_{1/2}|0\rangle=(|0\rangle+|1\rangle)/\sqrt{2}\tag{5}\\
    \hat{S}_{-1/2}|0\rangle=(|-1\rangle+|0\rangle)/\sqrt{2}\tag{6}\\
    \hat{S}_{1/2}|-1\rangle=(|-1\rangle+|0\rangle)/\sqrt{2}\tag{7}\\
    \hat{S}_{-1/2}|1\rangle=(|0\rangle+|1\rangle)/\sqrt{2}\tag{8}
\end{align*}

\par
\begin{figure}
    \centering
    \includegraphics[width=0.5\linewidth]{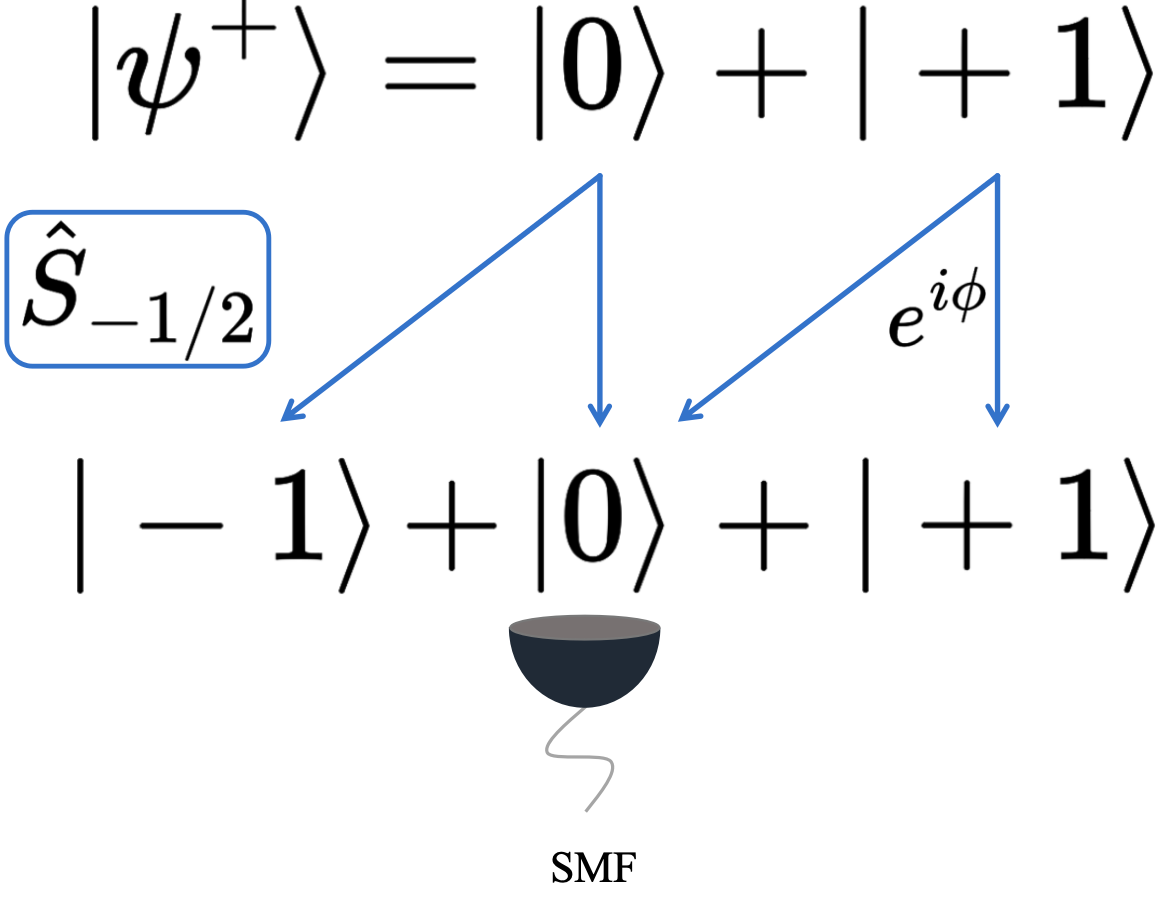}
    \caption{Schematic diagram of OAM label erasure. The translation phase plate $\hat{S}_{-1/2}$ operates on the input state, which is a superposition of $|0\rangle$ and $|+1\rangle$. As a result, some of the $|0\rangle$ component remains unchanged, while some of the components $|+1\rangle$ convert to $|0\rangle$. The final output state is a coherent superposition of $|0\rangle$ with a phase factor that depends on $\phi$.}
    \label{Fig2}
\end{figure}

 By applying this principle, the information from the OAM photon path can be erased. As illustrated in Figure \ref{Fig2}, an optical field at a superposition of $|0\rangle$ and $|1\rangle$ undergoes translation (i.e. deviation from the centre) after passing through a shifted SPP, where its basis states $|0\rangle$ and $|1\rangle$ are affected by the SPP. The base vector $|0\rangle$ is converted into a superposition of $|-1\rangle$ and $|0\rangle$ (Eq.6), while the base vector $|1\rangle$ is converted into a superposition of $|0\rangle$ and $|1\rangle$ (Eq.8). A single photon in a superposition state of $|0\rangle$ and $|1\rangle$ undergoes a transformation when its two state components are twisted by the shifted SPP. This results in a superposition state of $|-1\rangle$, $|0\rangle$ and $|1\rangle$, with the basis $|0\rangle$ acquiring a relative phase factor $e^{i\phi}$ from the MZI. The SPP induces a transformation of the superposition state, which can be expressed as 
\begin{align*}
       \hat{S}_{1/2}|\psi_{P3}\rangle&=\frac{1}{\sqrt{2}}(\hat{S}_{1/2}|0\rangle+e^{i\phi}\hat{S}_{1/2}|-1\rangle)\\
    &=\frac{1}{2}e^{i\phi}|-1\rangle+\frac{1}{2}(1+e^{i\phi})|0\rangle+\frac{1}{2}|+1\rangle\tag{9}\\
    \hat{S}_{-1/2}|\psi_{P4}\rangle&=\frac{1}{\sqrt{2}}(\hat{S}_{-1/2}|0\rangle+e^{i\phi}\hat{S}_{-1/2}|1\rangle)\\
    &=\frac{1}{2}|-1\rangle-\frac{1}{2}(1+e^{i\phi})|0\rangle+\frac{1}{2}e^{i\phi}|+1\rangle\tag{10}
\end{align*}

From Eqs. (9) and (10) we can see that both paths contain the same state except for the phase term. In this case, the path information is erased and the photon exhibits wave nature. The probability of observing a photon in the state $|0\rangle$ for $|\psi_{P3}\rangle$ is $|\langle 0|\hat{S}_{1/2}|\psi^+\rangle|^2=(1-\cos{\phi})/2$. Adjusting $\phi$ will allow for the observation of interference fringes of a single photon, i.e. the photon will show intrinsic volatility after exiting the MZI and being extinguished by the SPP.

\section{Experimental setup}
\begin{figure}
    \centering
    \includegraphics[width=0.75\linewidth]{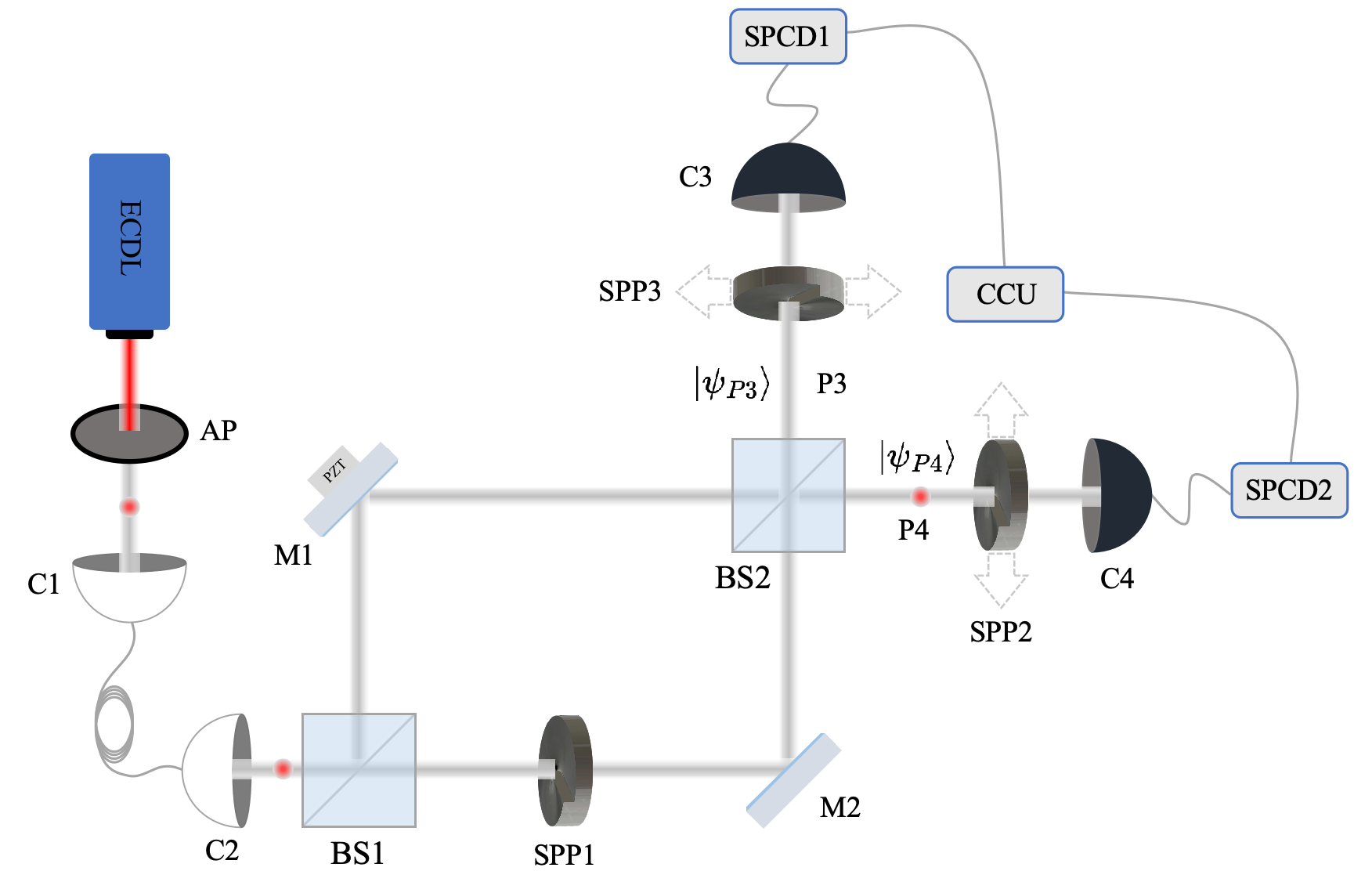}
    \caption{Schematic diagram of the experimental system. ECDL, External-Cavity Diode Laser; AP, Attenuator Plate; C, Coupler; BS, Beam Splitter; M, Mirror; A, Arm of the interferometer; PZT, Piezoelectric Transducer; SPP, Spiral Phase Plate; SPCD, Single Photon Count Detector; CCU, Coincidence Count Unit. A Mach-Zender interferometer is constructed by M1, M2, BS1, and BS2. Photons are detected and correlated by C3 and C4 and their corresponding devices on the two outputs of BS2.}
     \label{Fig3}
\end{figure}

The experimental system is shown in Fig.\ref{Fig3}. A 795 nm external-cavity diode laser (ECDL) is attenuated to the single-photon level ($1\times10^4$ photons/second) through an attenuator plate (AP) and coupled into single-mode fibre (SMF). Then this beam enters an MZI, which consists of two beam splitters (BS1 and BS2) and two mirrors (M1 and M2). The two arms of the MZI are labelled A1 and A2. A 1-th order spiral phase plate (SPP) was inserted in A2, which transforms the Gaussian mode $|0\rangle$ into the OAM state $|1\rangle$. A piezoelectric lead zirconate titanate (PZT) attached to M1 is controlled by a function generator. The function generator applies voltage signals to the PZT to precisely control the inclination angle of M1, thus adjusting the relative phase $\phi$ between A1 and A2. Photons passing through A1 and A2 are coherently superposed at BS2.

\begin{figure}
    \centering
    \includegraphics[width=0.75\linewidth]{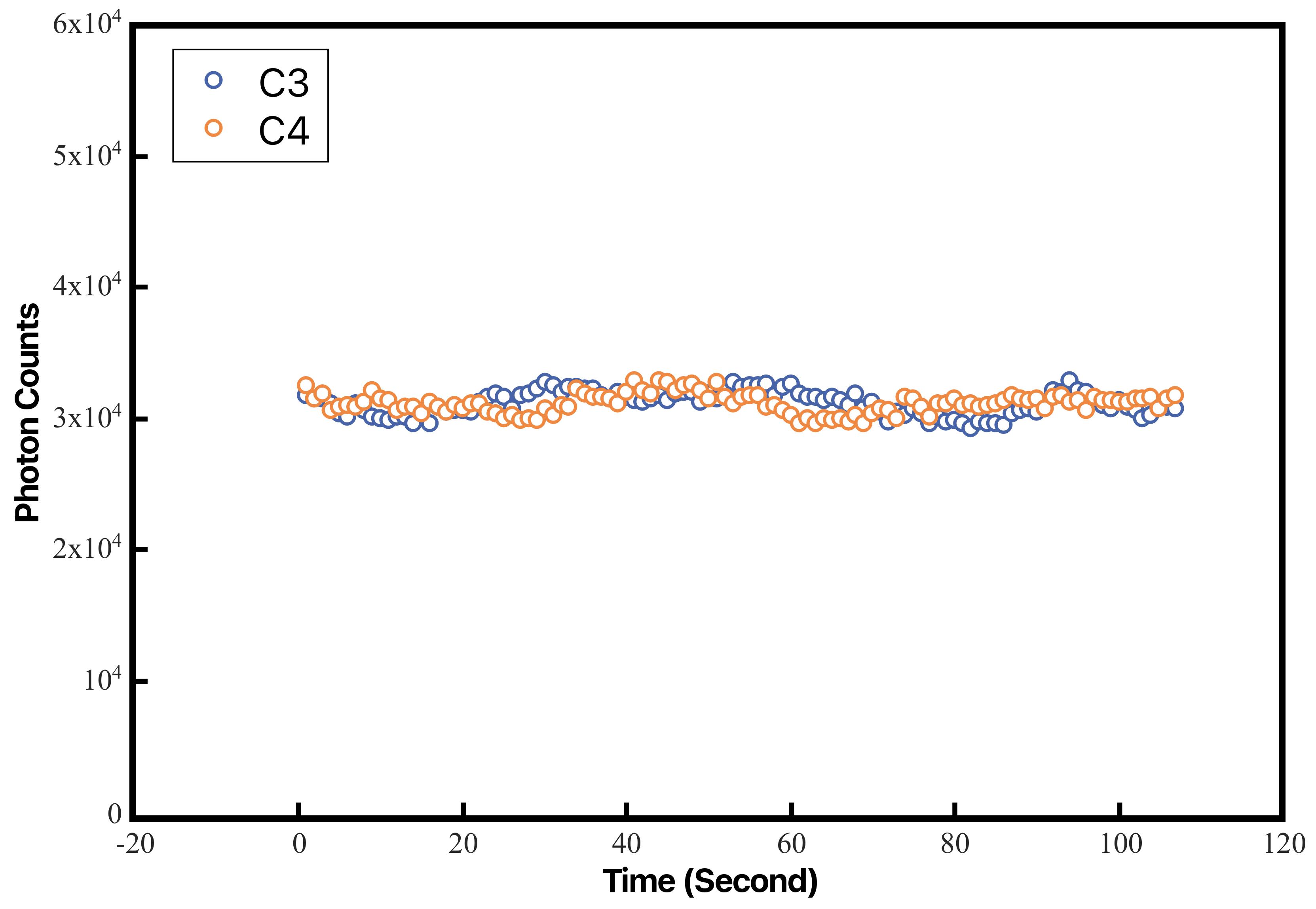}
    \caption{The number of photons detected by C3 and C4 depends on the relative phase $\phi$ of the light field projected onto the state $|0\rangle$. This figure shows the experimental data for the photon counts at C3 (blue dots) and C4 (orange dots) as a function of $\phi$. The relative phase $\phi$ is controlled by a piezoelectric transducer (PZT) that adjusts the optical path difference between the two arms of the Mach-Zehnder interferometer (MZI). The data demonstrate the correlation between the photon detection and the phase modulation.}
     \label{Fig4}
\end{figure}

Considering the inversion of the OAM mode caused by reflection ($|l\rangle$ becomes $|-l\rangle$ after reflection), the quantum states of the optical field in P3 and P4 are, respectively, $|\psi_{P3}\rangle=(e^{i\pi}|0\rangle+e^{i\phi}|-1\rangle)/\sqrt{2}$ and $|\psi_{P4}\rangle=(|0\rangle+e^{i(\phi+\pi)}|1\rangle)/\sqrt{2}$. The photons in the state $|0\rangle$ possess a flat equiphase surface, whereas those in the state $|1\rangle$ exhibit a helical equiphase surface, with the helical direction being opposite for $|-1\rangle$. If a photon passes through A1, it can be inferred by detecting a flat equiphase surface outside the MZI. If a photon is detected with a helically equiphase surface, this implies that it has passed through A2.
Photons are directed towards couplers C3 and C4, which are connected to SMF, when SPP2 and SPP3 are not inserted in the optical path. The detection device only allows the state $|0\rangle$ to enter at this operation, which corresponds to the projection measurement of the photons projected to the $|0\rangle$. Coincidence counting is achieved using single photon count detector(SPCD), while SPCD1 and SPCD2 are connected to the CCU. The coincidence counting rate was determined to be $1.25\times10^{-4}$ per counting time, indicating the presence of only one photon in the optical path during each counting time. Multiphoton interference is ruled out because the average distance between photons is approximately $3\times10^4$ metres. Figure \ref{Fig4} demonstrates the relationship between photon count and the relative phase phi of MZI for the projection measurement of $|0\rangle$ states.
\par Applying a triangular wave voltage with a frequency of 200mHz to PZT through a function generator will accurately change the inclination of M1, thereby linearly modulating the relative phase of the two arms of MZI. As shown in Figure \ref{Fig4}， the overall number of photons responded to by SPCD1 and SPCD2 is unaffected throughout the phase-scanning period, which means that if we can determine that the photon is passing through A1, the photon has particle properties and the light intensity does not depend on the phase $\phi$ of MZI. The constant number of photons detected by SPCD1 and SPCD2 throughout the scanning interval demonstrates the particle-like behaviour of a photon passing through A1, rendering the light intensity independent of the phase $\phi$ of MZI. The small variation is caused mainly by the unavoidable disturbance in the environment of the experimental system and the natural distribution of the coherent state light field under the representation of the Fock state\cite{RN445} $\sum_{n=0}^\infty C_n|n\rangle$$(C_1\approx1)$.
\par
\begin{figure}
    \centering
    \includegraphics[width=0.75\linewidth]{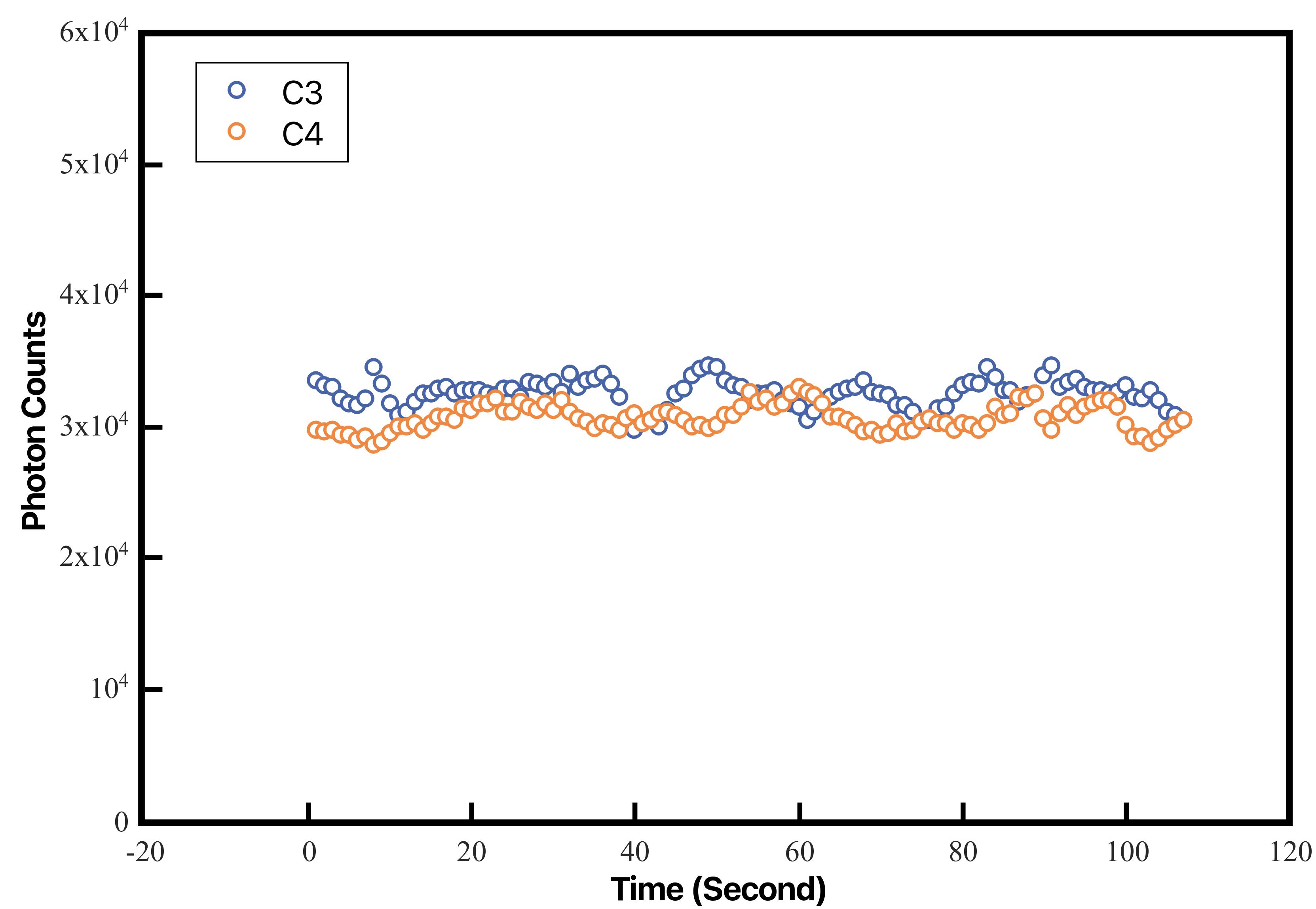}
    \caption{This figure shows the variation of the number of photons detected by C3 and C4 with the relative phase $\phi$ of the light field projected onto the $|1\rangle$ state. The blue and red dots in the figure represent the photon counts of C3 and C4, respectively. The figure clearly demonstrates the dependence of photon counts on the relative phase $\phi$, which is controlled by a PZT adjusting the optical path difference between the two arms of the MZI.}
     \label{Fig5}
\end{figure}

 To ensure proper photon transmission, the precise adjustment of the positions of SPP2 ($\hat{S}$) in the reverse direction and SPP3 ($\hat{S}^\dagger$) in the forward direction is crucial. The photons should pass through them perpendicularly at their centres. At P3, which includes SPP3, C3, and its SMF, a projection measurement is applied to the photon state $|-1\rangle$. Similarly, at P4, a projection measurement is applied to the state $|1\rangle$. Because $|\psi_{P3}\rangle=(e^{i\pi}|0\rangle+e^{i\phi}|-1\rangle)/\sqrt{2}$ and $|\psi_{P4}\rangle=(|0\rangle+e^{i(\phi+\pi)}|1\rangle)/\sqrt{2}$, the detection of a photon by an SPCD indicates that it has passed through A2. The photon count varies with the relative phase $\phi$ in the projection measurements on $|1\rangle$ and $|-1\rangle$, as illustrated in Figure \ref{Fig5},
when an SPCD detects a photon that has passed through A2, similar to the projection of $|0\rangle$, the photon exhibits a particle-like behaviour. The total number of photons remains approximately constant regardless of the phase $\phi$. The number of photons fluctuates slightly more than in Figure \ref{Fig4} due to the reduced robustness when SPP1 is connected in series with SPP2 or SPP3 on the optical path. Strict coaxial alignment at the centers is required for the photon to be perfectly projected to $|1\rangle$ or $|-1\rangle$, which is only possible in the ideal case when the centers of the two SPPs are perfectly coaxial and parallel. In actual experimental setups, the photon will be projected to $\frac{c_0|0\rangle+c_1|1\rangle}{\sqrt{2}}$ or $\frac{c_0|0\rangle+c_1|-1\rangle}{\sqrt{2}} (c_1\gg c_0)$. Projecting to $|1\rangle$ or $ |-1\rangle$ erases information about the phase structure of a few photons, leading to the phenomenon of wavelike behaviour.
\par If SPP2 (SPP3) is properly shifted to allow the transformation of a photon in the state $|0\rangle$ into the state $\frac{|0\rangle+|1\rangle}{\sqrt{2}}$ ($\frac{|0\rangle+|-1\rangle}{\sqrt{2}}$), then for a photon in the state $|\psi_{P4}\rangle$ ($|\psi_{P3}\rangle$) the following relationship holds:
\begin{align*}
       |\psi_{P4}^{-1/2}\rangle=\hat{S}_{-1/2}|\psi_{P4}\rangle&=\frac{1}{\sqrt{2}}(\hat{S}_{-1/2}|0\rangle-e^{i\phi}\hat{S}_{-1/2}|+1\rangle)\\
    &=\frac{1}{2}|-1\rangle+\frac{1}{2}(1-e^{i\phi})|0\rangle-\frac{1}{2}e^{i\phi}|+1\rangle\tag{9}\\
      |\psi_{P3}^{1/2}\rangle=\hat{S}_{1/2}|\psi_{P3}\rangle&=\frac{1}{\sqrt{2}}(-\hat{S}_{1/2}|0\rangle+e^{i\phi}\hat{S}_{1/2}|-1\rangle)\\
    &=\frac{1}{2}e^{i\phi}|-1\rangle-\frac{1}{2}(1+e^{i\phi})|0\rangle-\frac{1}{2}|+1\rangle\tag{10}
\end{align*}
Combining SMF to project $ |\psi_{P4}^{-1/2}\rangle$ and $|\psi_{P3}^{1/2}\rangle$ onto $|0\rangle$ can, under ideal conditions, completely erase their phase structure and path information. If SPCD responds and it is impossible to determine which path the photon has taken from the phase structure information, the photon will exhibit wave-like behaviour. The experimental results are shown in Figure 6:
\begin{figure}
    \centering
    \includegraphics[width=0.75\linewidth]{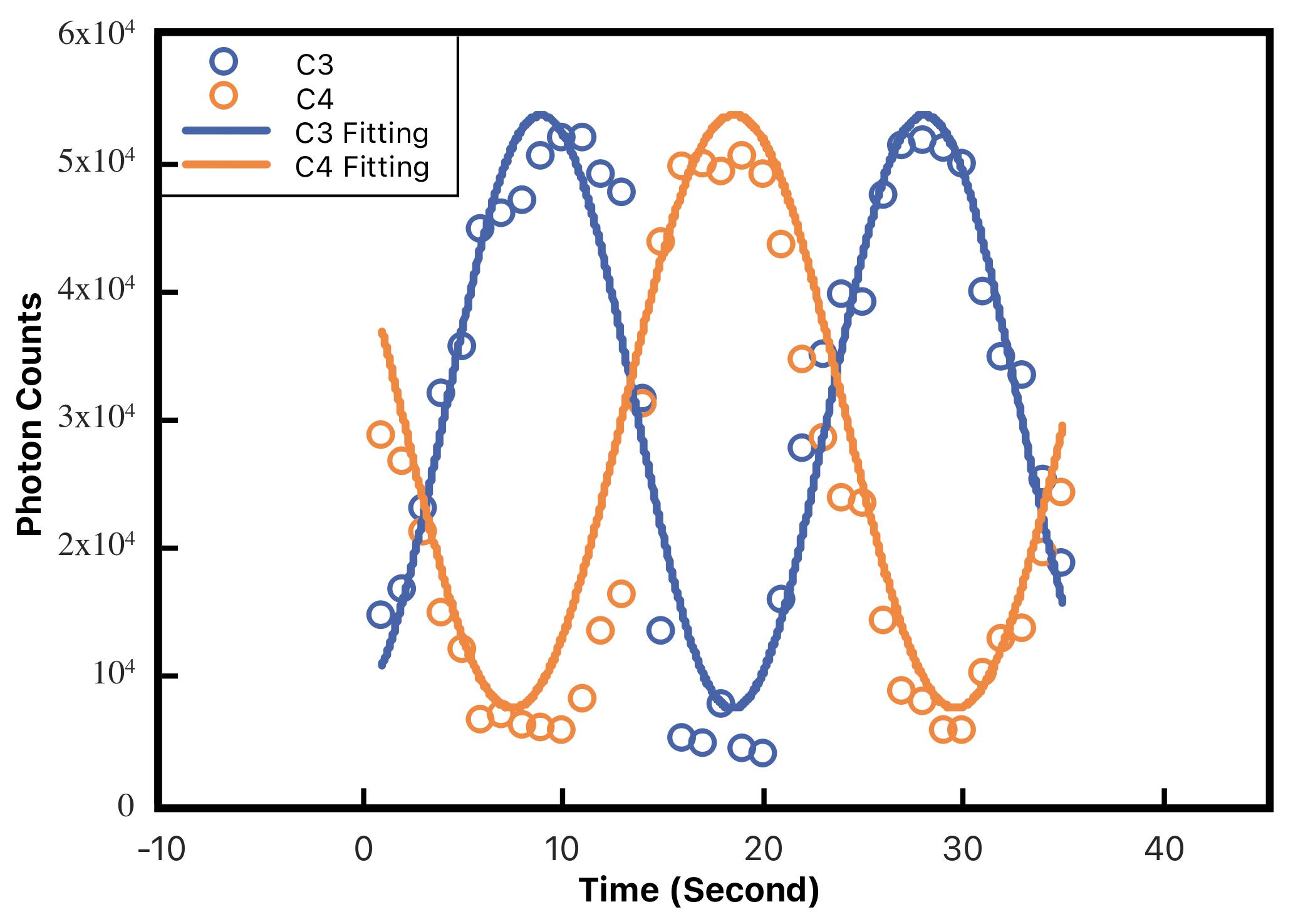} 
    \caption{Projection of the state $|0\rangle$ after erasing path information. The curve generated by fitting the data is consistent with the conclusion of the theoretical part, which is a sine curve dependent on $\phi$.}
     \label{Fig6}
\end{figure}
The number of photons exhibited clear wave-like fluctuations, and interference fringes appeared. The photon counts were fitted to a cosine function, and the relationship between the photon count and the phase $\phi$ was obtained as $N=A\cos{\phi}+C_0$, which is consistent with the theoretical prediction: $|\langle 0|\hat{S}_{1/2}|\psi^+\rangle|^2=(1-\cos{\phi})/2$. The visibility of the interference pattern can be quantified by equation: $V=(I_{max}-I_{min})/(I_{max}+I_{min})=84.35\% \pm1.7\%$, where $I_{min}$ and $I_{max}$ represent the minimum and maximum values of the photon counting rate with respect to the phase $\phi$.

\section{Discussion}

\par The complementarity between path information and fringe visibility represents a fundamental aspect of the quantum eraser. In order to further study the quantum erasing effect based on the phase structure, we conducted a theoretical analysis of the scheme in which the first-order OAM state generates internal state interference in MZI. This scheme was then realised experimentally. When the photon phase structure is erased, the final fringe contrast can reach $V_{max}=84.35\% \pm1.7\%$. In this work, for study the phase-erasure, we construct a concrete OAM channel that differs from the previous abstract channel\cite{Nape2017}，thus complements important previous work. This allows us to include the structure information of the equiphase plane of the light field in the quantum erasure.
\par In considering the evolution process of quantum systems, it is important to recognise that the picture of quantum physics should be viewed as a wholeness, rather than as a single solid particle. In an interferometer, even the wave packet of a single photon can be divided into two parts, each passing through a different path. In the experiment, the phase structure of the photon wave packet is modulated by inserting and adjusting SPP at a specific position in MZI. Subsequently, the phase structure is erased, resulting in internal state interference within the sigle photon wave packet. In summary: (1) This result serves to further expand and consolidate our quantum physical picture. (2) This paper presents methods for internal state interference that have potential value in quantum optics, such as atomic internal state regulation\cite{RN362}.

\section{Funding}
This work is supported by the National Natural Science Foundation of China (NSFC) (12104358,
12104361, and 12304406) and the Shaanxi Fundamental Science Research Project for Mathematics
and Physics (22JSZ004).

\bibliographystyle{unsrt}
  \bibliography{reference}

\end{CJK}
\end{document}